\documentclass[aps,twocolumn,
floats,prl,nofootinbib,superscriptaddress,10pt]{revtex4-1}
\usepackage{float}
\usepackage[utf8]{inputenc}

\usepackage{graphicx,amsmath,amsfonts,amssymb,slashed}
\usepackage{bbold,wasysym}
\usepackage{enumitem}

\usepackage{amsmath,amsfonts,bbm,euscript,hyperref}

\usepackage{xcolor}
\usepackage{color}
\usepackage{hyperref}

\hypersetup{colorlinks, citecolor=blue, linkcolor=red, urlcolor=blue}

\usepackage{graphicx,capt-of}

\newcounter{subfloat}

\newcounter{subfloat2}

\newcommand{\nco}{\newcommand}
\nco{\beq}{\begin{equation}} \nco{\eeq}{\end{equation}}
\nco{\beqa}{\begin{eqnarray}} \nco{\eeqa}{\end{eqnarray}}
\def\be{\begin{equation}}
\def\ee{\end{equation}}    
\def\baray{\begin{eqnarray}}
\def\earay{\end{eqnarray}}

\nco{\lra}{\leftrightarrow}

\def\pls{{\rm pulsar}}

\nco{\sss}{\scriptscriptstyle} \nco{\dphi}{\varphi}
\nco{\lsim}{\mbox{\raisebox{-.6ex}{~$\stackrel{<}{\sim}$~}}}
\nco{\gsim}{\mbox{\raisebox{-.6ex}{~$\stackrel{>}{\sim}$~}}}

\def\IK{\relax{\rm I\kern-.20em K}}
\def\IM{\relax{\rm I\kern-.20em M}}

\def\lsim{\mbox{\raisebox{-.6ex}{~$\stackrel{<}{\sim}$~}}}
\def\gsim{\mbox{\raisebox{-.6ex}{~$\stackrel{>}{\sim}$~}}}
\def\sss{\scriptscriptstyle}

\def\ga{\mathrel{\raise.3ex\hbox{$>$\kern-.75em\lower1ex\hbox{$\sim$}}}}
\def\la{\mathrel{\raise.3ex\hbox{$<$\kern-.75em\lower1ex\hbox{$\sim$}}}}

\usepackage{amsmath}
\usepackage{lipsum}

\begin{document}

\title{Probing ultralight  scalar, vector and tensor dark matter  with pulsar timing arrays}
\author{Caner \"Unal } 
\email{unalx005@umn.edu}
\affiliation{Department of Physics, Ben-Gurion University of the Negev, Be'er Sheva 84105, Israel}
\affiliation{Feza Gursey Institute, Bogazici University, Kandilli, 34684, Istanbul, Turkey}

 \author{Federico R. Urban} 
 \email{federico.urban@fzu.cz}
\affiliation{CEICO, Institute of Physics of the Czech Academy of Sciences, 182 21 Prague 8, Czech Republic} %Na Slovance 1999/2, 182 21 Prague 8, Czech Republic}

\author{Ely D. Kovetz} 
\email{kovetz@bgu.ac.il}
\affiliation{Department of Physics, Ben-Gurion University of the Negev, Be'er Sheva 84105, Israel}

\begin{abstract} 
Pulsar timing arrays (PTAs) are sensitive to oscillations in the gravitational potential along the line-of-sight due to ultralight particle pressure. 
We calculate the probing power of PTAs for ultralight bosons across all frequencies, from those larger than the inverse observation time  
to  those smaller than the inverse distance to the pulsar.  We show that  since the signal amplitude grows comparably  to the degradation in PTA sensitivity at frequencies smaller than inverse observation time,  the discovery potential can be extended towards lower masses by over three decades, maintaining high precision. We demonstrate that, in the mass range $10^{-26} -10^{-23}$ eV, existing 15-year PTA data can robustly detect or rule out an ultralight component down to $O(1 - 10)\%$ of the total dark matter. Non-detection, together with other bounds in different mass ranges, will imply that ultralight scalar/axion can comprise at most $1-10\%$ of dark matter in the $10^{-30}\!-\!10^{-17}$ eV range. With 30 years of observation, current PTAs can extend the reach down to  $0.1-1 \%$, while next-generation PTAs such as SKA can attain the $0.01-0.1\%$ precision. We generalize and derive predictions for ultralight spin-1 vector (i.e.\ dark photon) and spin-2 tensor dark  components. 
\end{abstract}

\maketitle

Ultralight particles are intriguing dark matter candidates, motivated by high energy theories~\cite{Svrcek:2006yi}. As they have de-Broglie wavelengths comparably to galactic or even larger cluster scales, depending on their mass, they would leave marks on cosmological and astrophysical observables~\cite{Arvanitaki:2009fg}. One such effect comes from the pressure of the ultralight field which could cause the energy-momentum tensor and equivalently spacetime to fluctuate in a monochromatic way~\cite{Khmelnitsky:2013lxt}. These fluctuations would modify the observed arrival time of pulses from pulsars and can be used to extract the properties, e.g.\ mass and energy density, of ultralight scalar/axion-like (spin-0), vector (dark photon/spin-1) and tensor (spin-2) bosons, even if they form a fraction~\cite{Blum:2021oxj,Ye:2021iwa} of dark matter.

Ultralight bosonic degrees of freedom have undergone intense exploration in recent years (see Ref.~\cite{Ferreira:2020fam} for a review). The mass parameter space has already been constrained at several distinct scales by a number of probes. Cosmic microwave background experiments (CMB)~\cite{Hlozek:2014lca,Poulin:2018dzj} combined with large-scale structure surveys~\cite{Lague:2021frh} have derived bounds on the extremely light range, roughly $10^{-32}\!-\!10^{-25}\,{\rm eV}$. The ultraviolet luminosity function and reionization constrain the $10^{-23}\!-\!10^{-22}\,{\rm eV}$ range~\cite{Bozek:2014uqa}. Lyman-$\alpha$ experiments reaching large wavenumbers of the matter power spectrum probe approximately the range $10^{-23}\!-\!10^{-20.5}\,{\rm eV}$~\cite{Kobayashi:2017jcf,Irsic:2017yje,Armengaud:2017nkf,Rogers:2020ltq,Rogers:2020cup}, and the shape of Eridanus-II the $10^{-20.5}-10^{-19} \, {\rm eV}$ range\cite{Marsh:2018zyw}. Galaxy rotation curves probe the $10^{-23.7}-10^{-22} \, {\rm eV}$ range~\cite{Bar:2021kti}. Secular variations in binary-pulsar orbital parameters can test the mass range $10^{-23}\!-\!10^{-18}\,{\rm eV}$~\cite{Blas:2019hxz,Armaleo:2019gil,LopezNacir:2018epg,Blas:2016ddr}. 
Finally, non-observation of superradiance in  rapidly spinning supermassive black holes (SMBHs) can explore the $10^{-20.5}\!-\!10^{-16.7}\,{\rm eV}$ range~\cite{Unal:2020jiy}. With all of the above, the mass range $10^{-26}\!-\!10^{-23}\,{\rm eV}$ still remains relatively unexplored.

In this {\it Letter}
, we show that  current  data from the International Pulsar-Timing Array (IPTA)~\cite{Manchester:2013ndt,Verbiest:2016vem}, a collaboration between  NANOGRAV~\cite{NANOGrav:2015aud}, PPTA~\cite{Manchester:2012za} and EPTA~\cite{Lentati:2015qwp} %and InPTA~\cite{Tarafdar:2022toa}, 
can be used to probe the ultralight boson energy density fraction ${\cal F}_{\rm ULDM}$ in the $10^{-26}\!-\!10^{-23}\,{\rm eV}$ mass range with about 1-10\% precision. The bounds remarkably extend to the smaller masses (lower frequencies) in this window, since the strength of the ultralight particle signal changes comparable to the PTA sensitivity curve. Therefore, smaller mass particles can be efficiently constrained by current IPTA data down to frequencies corresponding to $1/D_{\rm pulsar}\!\sim\! 1/{\rm kpc}$ (where $D_{\rm pulsar}$ is the distance to the pulsar, and we set the speed of light $c=1$).  Extending the  observation period up to 30 years can improve these constraints by one order of magnitude, and with future PTA experiments, such as SKA~\cite{Weltman:2018zrl}, by two more orders of magnitude. 
Combined with CMB/LSS, Lyman-$\alpha$ and SMBH superradiance constraints, the discovery potential of PTAs implies that  ultralight bosons can be probed continuously throughout the mass range $10^{-30}\!-\!10^{-17}$ eV with $\sim {\cal O}(1)\% $ precision (see \cite{Hlozek:2016lzm,Safarzadeh:2019sre,Dentler:2021zij,Flitter:2022pzf,Libanore:2022ntl,Sun:2021yra}). 

We start with the signature of ultralight dark matter in PTAs. Free scalar/axion-like degrees of freedom, $\phi$, with a canonical kinetic term have a Lagrangian density ${\cal L}= \frac{1}{2} (\partial_\mu \phi)^2  - \frac{1}{2}  m^2 \phi^2$. If such particles are non-relativistic, then the field configuration is given by a plane wave of nearly single frequency with corrections up to the kinetic term, which is about $10^{-6}$ times smaller compared to the rest mass.  If we neglect the expansion of the background,  $\phi \!=\!{\cal A}(x) \, \cos(mt + \beta)$. The energy density is nearly time independent, $\rho = \dot \phi^2/2+V\simeq \frac{1}{2} m^2 {\cal A}^2$, and the pressure oscillates with angular frequency twice the mass of the particle, i.e.  $p = \dot \phi^2/2- V\simeq \rho \cdot  \cos(2\pi f t + 2\beta)$, where $f$  is the oscillation frequency that can be computed using the corresponding mass as $2\pi f = w= 2m$, and {\color{red} $\beta$} is a phase. Hence, 
\begin{equation}
f=5 \cdot10^{-9}{\rm Hz} \left(\frac{m}{10^{-23}\, {\rm eV}} \right).
\label{eqfreqmass}
\end{equation}
Then, solving the 00 and ii components of Einstein's equations at first order, with the following metric perturbations: $\delta g_{00}=-2\Phi$ and $\delta g_{ij} = 2\psi \delta_{ij}$, one ends up with  $\psi = \psi_0(x)+ \frac{\pi}{2} G_N {\cal A}^2  \cos (2mt+ 2 \beta) = \psi_0(x)+ \pi G_N \rho /m^2  \cos (2mt+ 2 \beta)$, the second term being the oscillating piece defined as $\psi_c\!=\! \pi G_N \rho /m^2$. The time residual can be calculated via the fractional frequency shift as
\begin{equation}
\delta \Delta t = \int_{t_p}^{t}  \frac{\nu' -  {\bar \nu}}{{\bar \nu}} dt,
\label{eqresidualdefine}
\end{equation}
where $t$ and $t_p$ are the time at Earth and pulsar emission, and the fractional change in the frequency is given by 
\begin{equation}
\frac{\nu' - {\bar \nu}}{{\bar \nu}}  = \psi (x,t) - \psi (x_p,t_p)  - \int_{t_p}^{t} n_i \partial_i (\Phi + \psi) dt',
\label{eqnus}
\end{equation}
where the first term is the difference in the gravitational potential and the second term is the change of its gradient during propagation, which is suppressed by a factor $k/m$. 
Now, plugging Eq.~\eqref{eqnus} into Eq.~\eqref{eqresidualdefine}, we get the expression
\begin{eqnarray}
\delta \Delta t &=& \frac{\psi_c}{m}  \sin (m D_{\rm pulsar}+ \beta_e - \beta_p) \times  \nonumber\\
&& \times \cos (2mt - mD_{\rm pulsar}+ \beta_e + \beta_p).
\label{eqtimeresidual}
\end{eqnarray}
where $\beta_e$ and $\beta_p$ corresponds to phase values at the Earth and at the individual pulsar.  The typical amplitude for the residual oscillating  signal   is the root-mean-square (rms) path average which is given by\footnote{For simplicity, we assume that for cross-correlation of signals from two different pulsars, this factorizes as ${\cal P}(L_1){\cal P}(L_2)$ where $L_1$ and $L_2$ are the Earth-pulsar distances for pulsar 1 and 2.} 
\begin{equation}
\sqrt{ \langle ( \delta \Delta t )^2  \rangle } =  \sqrt{ \frac{1}{L}  \int_0^{L}  dl \, (\delta \Delta t)^2 } = {\cal P} \cdot \psi_c/ m,
\end{equation}
where $L\equiv D_\pls$ and ${\cal P} $  is defined as
\begin{eqnarray}
\!\!\!\! {\cal P}  
%&\equiv& \sqrt{ \frac{1}{L} \int_0^{L} \sin^2( m\, x) dx  }
= \frac{1}{\sqrt{2}} \left(1- \frac{ \sin( 2\, m \, L)}{ 2\, m \,L} \right)^\frac{1}{2}\!\!\,\,  %\left(1- \frac{ \sin( 2\, m \, T_{obs})}{ 2\, m \,T_{obs}} \right)^\frac{1}{2}
\end{eqnarray}
In the case where $m \cdot D_\pls > 1$ %and $m \cdot T_{obs} > 1$,
we have ${\cal P}\simeq1$. However, in the limit $m \cdot D_\pls\ll1$, %or $m \cdot T_{obs} \ll1$, 
we have extra suppression factor in the signal, of approximately $m \, D_\pls$. % and $m \, T_{obs}$, respectively.

The oscillations in the energy-momentum tensor will cause oscillations in the gravitational potential which correspond to an equivalent characteristic strain $h_c =2\sqrt{3} \psi_c$, stressing that this is not GW signal.  A similar analysis can be done for vector and tensor fields. For the massive spin-1 case, i.e.\ a massive dark photon, the field has 2 transverse and 1 longitudinal polarizations. For the spin-2 case, i.e.\ tensor, the field has 1 scalar, 2 transverse and 2 transverse-traceless polarizations (for subluminal gravitational waves \cite{Qin:2020hfy}).  These polarizations are of similar order-of-magnitude in strength. Although there will be accompanying polarizations, the distinctive modes will be transverse polarization for a vector field and transverse-traceless polarization for a tensor field, and so we will conduct our analysis for the detectability of these specific polarizations. The signal from a single pulsar corresponding to the scalar~\cite{Khmelnitsky:2013lxt}, vector~\cite{Nomura:2019cvc} and tensor~\cite{Armaleo:2020yml} dark matter scenarios can be estimated as 
\begin{eqnarray}
&&\!\!\!\! \!\!\!\! h_{c, \, {\rm scalar}} \simeq 2\cdot 10^{-15}  \cdot {\cal P}  \cdot  {\cal F}_{\rm ULDM} \left(  \frac{ 5 {\rm nHz}}{f}   \right)^2   \,       
 \nonumber\\
&&\!\!\!\! \!\!\!\!h_{c, \, {\rm vector}} \simeq 6 \cdot 10^{-15}  \cdot {\cal P}  \cdot    {\cal F}_{\rm ULDM}  \left(  \frac{ 5 {\rm nHz}}{f}   \right)^2     \,   \nonumber \\ 
&&\!\!\!\! \!\!\!\! h_{c, {\rm tensor}, \, \alpha} \simeq 5 \cdot 10^{-15}   \cdot {\cal P}  \cdot      {\cal F}^{1/2}_{\rm ULDM} \,\,
 \left(  \frac{ 5 {\rm nHz}}{f}  \right)   \left(\frac{\alpha}{10^{-7}}\right)  \;\;\;
\label{eqsignal}
\end{eqnarray}
where ${\cal F}_{\rm ULDM} \equiv (\frac{\rho_{\rm ULDM}}{0.4 \, GeV/cm^3} )$ is the fraction of ultralight dark matter (ULDM) in the local Universe and $\alpha$ is the spin-2 universal coupling to matter. 

\begin{figure}[h!]
\includegraphics[width=0.49\textwidth,angle=0,scale=1]{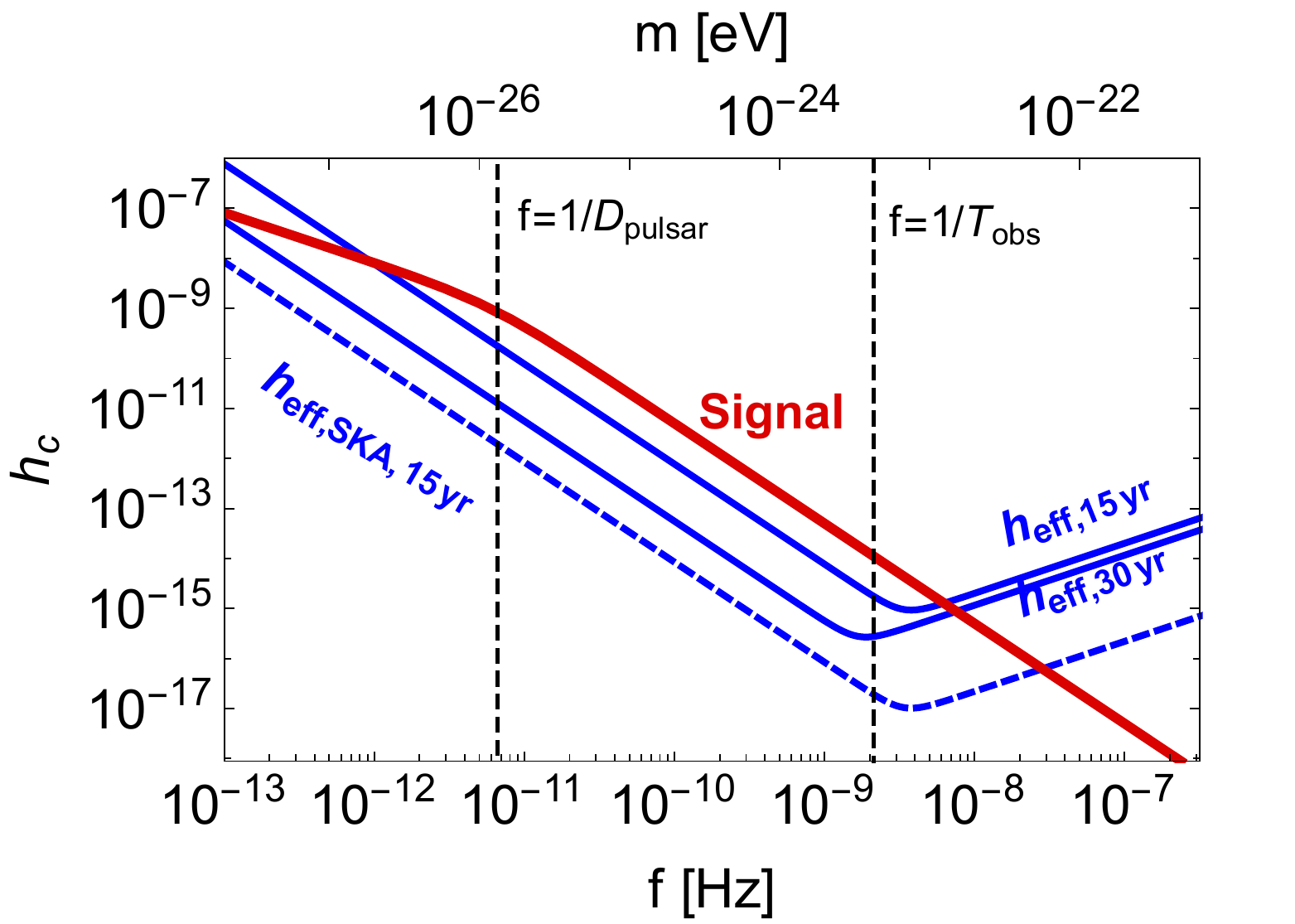}
\vspace{-0.2in}
\caption{Characteristic strain, $h_c$, for the signal in the case of scalar ultralight particles with ${\cal F}_{\rm ULDM} = 1$, and for the noise with current PTAs (assuming 15 and 30 years of observation time \cite{Chen:2021rqp}, for 60 and 90 pulsars) and with SKA (15 years with 5K pulsars), respectively. The effective noise curve is defined here as $h_{\rm eff} \equiv h_{\rm noise} (2  {\cal R}_{\beta}^2)^{-1/4}$. The signal, the PTA sensitivity and the detectability in three
regimes separated by vertical dashed lines are discussed in detail in the text. Note that the signal grows comparable
to the the noise in the regime $1/D_{\rm pulsar} < f < 1/T_{\rm obs}$, allowing for precision probing, and slower for $f > 1/T_{\rm obs}$ and $f < 1/D_{\rm pulsar}$.
}
\vspace{-0.15in}
\label{signalvssensitivity}
\end{figure}

A number of comments are in order. First, the signal suppression from propagation is important, since with the corresponding modification the signal strain grows more slowly than the red-noise strain and as a result the signal-to-noise diminishes for frequencies smaller than inverse pulsar distance, i.e.\ $f<1/D_\pls$, see Fig. \ref{signalvssensitivity}.

\begin{figure}[h!]
\centering 
\includegraphics[width=0.49\textwidth,angle=0,scale=1]{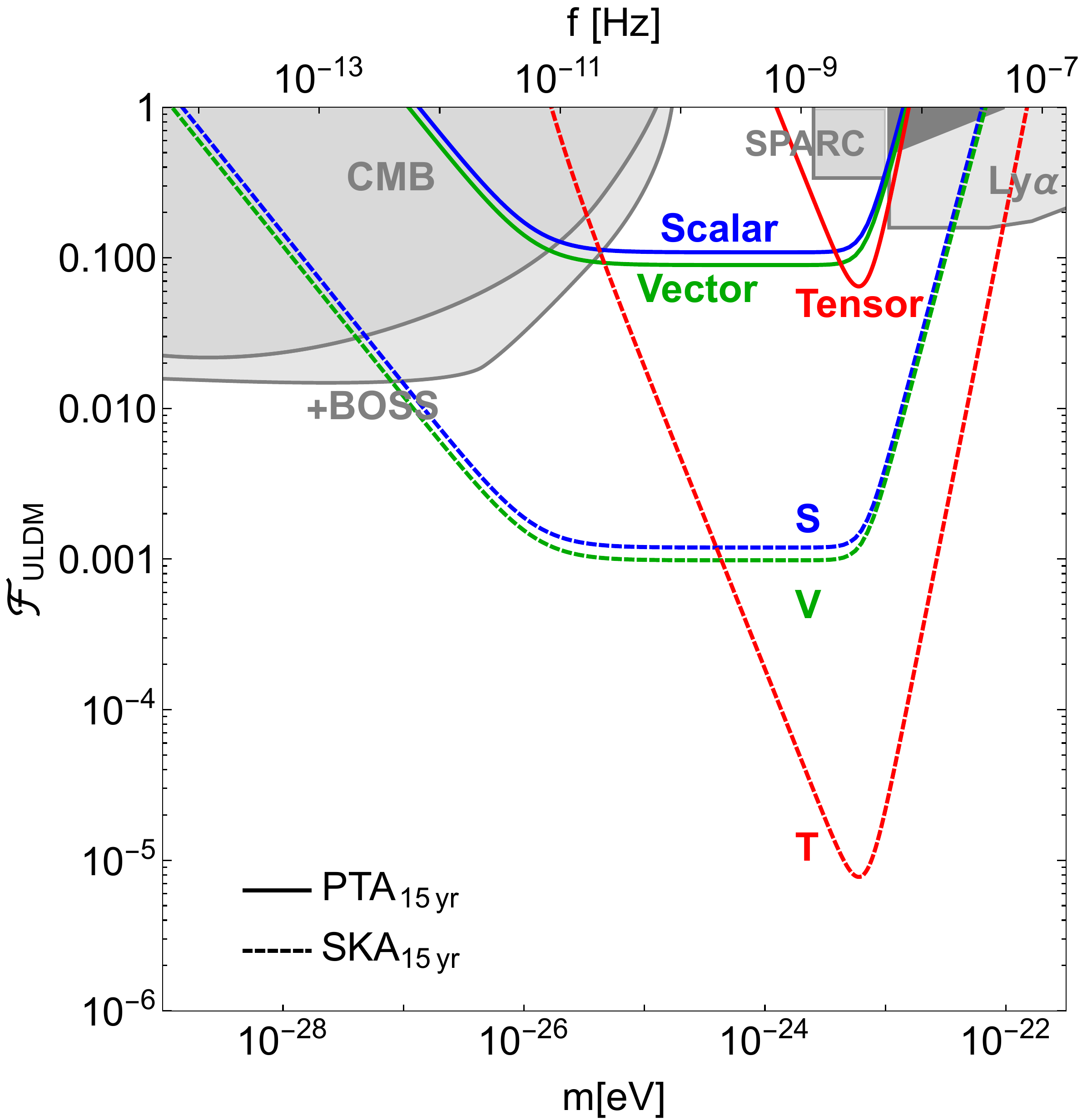}
\caption{Zoomed in for scalar, vector (dark photon) and tensor (spin-2) (setting $\alpha\!=\!10^{-7}, N_p^{PTA}=60, N_p^{SKA}=5K$, and 15 years of observation time for both).}
\label{figscalarvectorspin2zoom}
\end{figure}

Secondly, the tensor case is special in the sense that the only known viable massive spin-2 theory, bigravity, includes a universal direct coupling term to matter with strength $\alpha$ that cannot be tuned away. Because of this, the signal from the massive tensor field is proportional to $f^{-1}$ instead of $f^{-2}$ as in the scalar and vector cases~\cite{Armaleo:2020yml}. Therefore, in the tensor case the constraint is effectively on the combination $\alpha^2{\cal F}_{\rm ULDM}$ \footnote{This harder scaling is what makes it possible to potentially detect this kind of signal also at the much higher frequencies probed by gravitational wave interferometers~\cite{Armaleo:2020efr,Pierce:2018xmy,LIGOScientificCollaborationVirgoCollaboration:2021eyz}, see Fig.~\ref{figscalarvectorspin2zoom}.}$^{,}\!\!$
\footnote{We note that the same \(f^{-1}\) scaling also applies when spin-0 and spin-1 ULDM are directly coupled to matter
\begin{align}
\!\!\!\! h_{c, \,{\rm  scalar}, \,  \Lambda} &\simeq 3\cdot 10^{-18} \left(\frac{{\rm v}}{10^{-3}}\right) \left(\frac{10^{-26}\,GeV}{\Lambda}\right)   {\cal F}_{\rm ULDM} \,   \frac{ 5 {\rm nHz}}{f} \nonumber\\
\!\!\!\!h_{c, \, {\rm vector}, \, {\rm gc}} &\simeq 3.2\cdot 10^{-13} \left(\frac{{\rm g}}{10^{-24}}\right)  \, {\cal F}_{\rm ULDM}  \,  \frac{ 5 {\rm nHz}}{f}  
\end{align}
where $\Lambda$ is a cutoff, and "g" denotes the coupling constant.
The interaction Lagrangian that defines the coupling constants quoted above is
%%
%\begin{eqnarray}
%&&{\cal L} = M_\oplus\left(1+\frac{\phi(t)}{\Lambda}\right) \, \left(1+\frac{v^2}{2}\right) \nonumber \\ &&+ q v_i A^i(t) + \frac{\alpha M_\oplus}{2M_P} M_{ij}(t) v^i v^j \,,
%\end{eqnarray}
%%
%
\begin{equation}
{\cal L} = M_\oplus\left(1+\frac{\phi(t)}{\Lambda}\right) \, \left(1+\frac{v^2}{2}\right)+ q v_i A^i(t) + \frac{\alpha M_\oplus}{2M_P} M_{ij}(t) v^i v^j \,,
\end{equation}
where the spin-0, spin-1 and spin-2 fields are \(\phi(t)\), \(A_i(t)\) and \(M_{ij}(t)\), respectively; \(M_\oplus\) is the Earth's mass and \(v_i\) the velocity of the solar-system barycenter quasi-inertial frame with respect to the ULDM frame, \(M_P\) the reduced Planck mass, and \(q=g M_\oplus / m_n\) with \(m_n\) the mass of the neutron and \(g\) the \(B-L\) or \(B\) coupling constant.
}.

Having described the three different types of signals we are after, we now turn to a discussion of the sensitivity of current and future PTAs. Although all ultralight particles generate pressure which leads to perturbations in the arrival time of pulsar signals, the type of perturbations are distinct from each other. Accordingly, PTAs have different response functions for scalar, vector and tensor fields. 
The variables typically used in the literature to indicate the strength of the signal: the characteristic strain $h_c$, the power spectral density $S_h$ and the energy density $\Omega_h$, are connected to each other as follows
\begin{equation}
H_0^2 \,  \Omega (f) =   \frac{2\pi^2}{3} f^3 S_h(f)  =   \frac{2\pi^2}{3} f^2 h_c^2(f). 
\label{eqrelate}
\end{equation}
To estimate the detectability of the monochromatic signal, we start from the formula
\begin{eqnarray}
\!\!\!\! \!\!\!{\rm SNR}^2 = 2 \;  ( f \cdot T_{\rm obs} )^2    \sum_{I=1}^{N} \sum_{J>I}^{N}   \; r_{(\beta) \; IJ}^2\;   \left( \frac{S_{h,{\rm signal}}} {S_{h,{\rm noise}}}
 \right)^2,\,\,\,\,\,  
\label{eq:snrsquare}
\end{eqnarray}
where  $T_{\rm obs}$ is the observation time, and $I,J$ indicate  two pulsars in each pair correlation, and $r_{(\beta){IJ}}^2 \equiv \frac{1}{4\pi} \int d {\Omega} \; \chi^2_{IJ} $, $ \beta$=Scalar (S), Vector(V), Tensor(T), and $N$ is the total number of pulsars, assumed to have identical properties to derive an ensemble average. The explicit expressions for $\chi_{IJ}$, the correlation coefficient between each pair of pulsars separated by an angle $\zeta$, for  each $\beta$ are \cite{Jenet:2014bea}
\begin{eqnarray}
&&{\rm S} : \;   \chi_{IJ}=1, ~~~~~~~  {\rm V} : \chi_{IJ}=\frac{1}{3} \cos{\zeta} \nonumber \\
%&& \nonumber \\
&& {\rm T} : \chi_{IJ}=\frac{1}{2} - \frac{1}{4}\left( \frac{1-\cos{\zeta}}{2}\right) \left (1 + 6 \ln{\left( \frac{1-\cos{\zeta}}{2}\right)} \right)\, \nonumber 
 \end{eqnarray}
Defining ${\cal R}^2_{\beta}  \equiv  \sum_{I=1}^{N} \sum_{J>I}^{N} r^2_{(\beta) IJ}$, this yields
\begin{equation}
\!\!\!\!\!\! {\cal R}_{\rm S}^2= \frac{N \, (N-1)} {2 }, \;   {\cal R}_{\rm V}^2= \frac{N \, (N-1)} {2 \cdot 27 } , \;  {\cal R}_{\rm T}^2 = \frac{N \, (N-1)} {2 \cdot48}.
\label{eqptaspinresponse}
\end{equation}
The PTA sensitivity curve can be described as 
\begin{equation}
S_n = 12 \pi^2 f^2 \, \frac{{\cal N}}{{\cal T}} = h_n^2 / f
\end{equation}
where ${\cal T}= \left(\frac{(f\cdot T_{\rm obs})^3}{1+(f\cdot T_{\rm obs})^3}\right)^2$ is  the transmission function\footnote{
The transmission function in the limit $f<1/T_{\rm obs}$ accounts for the fact that fitting for the unknown pulsar period and period derivative is equivalent to keeping the transmission function fixed at 1 and expanding our signal up to cubic order, which then introduces the $f^3$ dependence to the signal.}, ${\cal N} = 2 \sigma^2 \Delta t + P_{r} \, f^{-\gamma}$ (white noise plus red noise which can be relevant for small frequencies). $S_n$  varies between the different frequency ranges (see Fig.~\ref{signalvssensitivity}):

 \begin{figure*}
\centering 
\includegraphics[width=0.8\textwidth,angle=0]{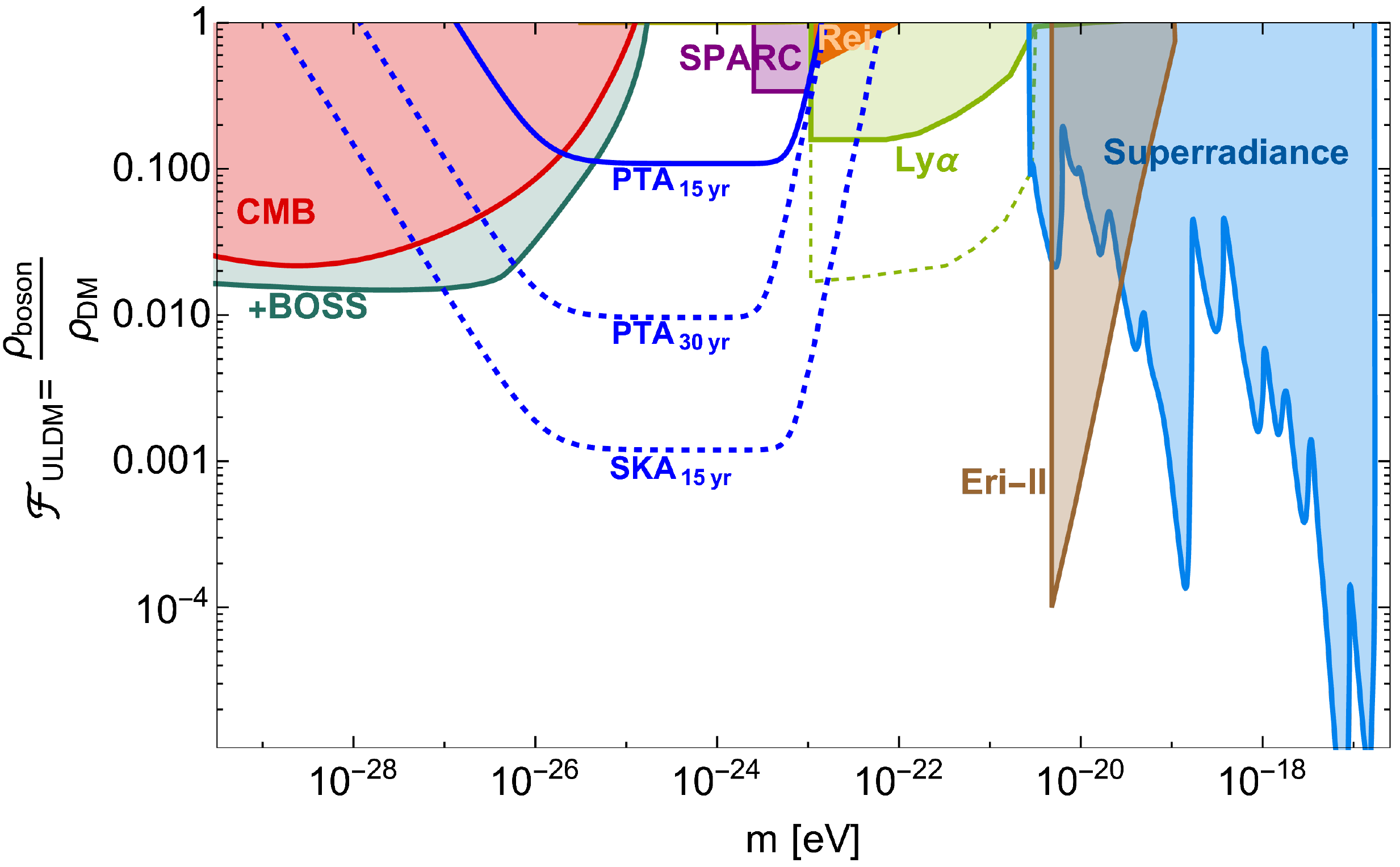}
\caption{
Bounds on the fraction of a scalar ultralight boson component out of the total dark matter.
We show existing constraints from CMB~\cite{Hlozek:2014lca,Poulin:2018dzj}; combined with BOSS~\cite{Lague:2021frh}; reionization~\cite{Bozek:2014uqa}; Lyman-$\alpha$~\cite{Kobayashi:2017jcf,Irsic:2017yje,Armengaud:2017nkf,Rogers:2020ltq,Rogers:2020cup}; Eridanus-II~\cite{Marsh:2018zyw};  galaxy rotation curves~\cite{Bar:2021kti};  
and SMBH superradiance~\cite{Unal:2020jiy}. Our forecasts for current PTA (future; 30 year PTA, and 15 year SKA) are shown in solid (dashed) blue.  %Upper (lower) curves correspond to conservative (optimal) red noise scaling, ie. $\gamma_r=1$ ($\gamma_r=0.75$).
}
\label{figbigpicture}
\vspace{-0.15in}
\end{figure*}

\begin{itemize}[leftmargin=*]
\item[] I) $f>1/T_{\rm obs}$  : The high frequency part of the sensitivity regime is controlled by white timing noise, whose expression is given as $S_n(f)=12\pi^2 f^2  (2 \Delta t \sigma^2) = h_n^2/f$. Here $\Delta t $ is the timing period, and $\sigma$ is the rms error in timing residuals. This frequency regime has scaling $h_n \propto f^{3/2}$, or equivalently $\Omega_n \propto f^{5}$ \cite{Thrane:2013oya,Moore:2014lga,Romano:2016dpx,Vigeland:2016nmm}.
\item[]II) $f<1/T_{\rm obs}$: Here $S_n$ is set by   three factors: (i) the transmission function, which accounts for the information absorbed by the timing model fit (we assume a quadratic spin-down model which fits for the pulsar phase offset, spin period and period derivative). In this regime, the transmission function scales as $f^{-6}$ and limits the detection capability ~\cite{Hazboun:2019vhv}. (ii) Measurement white noise, which is frequency-independent; (iii) Pulsar specific red noise \cite{Goncharov:2021oub}, which is more effective at low frequencies  \cite{Lam:2016iie} (see also \cite{NANOGrav:2020bcs,Goncharov:2020krd}). For simplicity, in making our forecast we will limit ourselves to a subset of pulsars for which, in the frequency regime we focus on, the red noise is subdominant compared to white noise. We assume that this holds for 1/4th of the total dataset (indeed, there are many such pulsars ~\cite{Cordes:2010fh,Lam:2016iie,NANOGrav:2020bcs,Goncharov:2020krd,NANOGrav:2020qll}).

When the frequency of the signal is less than $1/T_{obs}$, there are two ways to do the analysis:
i) One can keep the white noise fixed and expand the signal in a Taylor series. Then, the first two expansion terms are absorbed by the period and period derivative terms due to the lack of an independent measurement of these parameters, hence we are only left at the next order with the cubic term. Keeping the noise fixed, we therefore have an extra $m^3/T^3 \sim f^3/T^3$ suppression.
ii) One can keep the signal the same, but then the response will include again the ambiguity in period and period derivative, hence one can modify the definition of the noise via the transfer function \cite{Hazboun:2019vhv}. The strain transfer function is unity for frequencies larger than inverse observation time ($f > 1/T_{obs}$), on the other hand, and it is $f^{-3}$ at low frequencies ($f > 1/T_{obs}$). The transfer function for power spectral amplitude is quadratic in the strain, hence they have $f^{-6}$. Red noise makes the sensitivity worse on top of this white noise curve.
\end{itemize}

The noise curve for a generic single pulsar is parametrized as in Ref.~\cite{Unal:2020mts}, consistent with both simulated curves~\cite{Hazboun:2019vhv} and data~\cite{NANOGrav:2020qll}

\begin{eqnarray}
\!\!\!\!\!\! \!\!\!\!\!\! h_{c,{\rm noise}} &=& \sqrt{f\cdot S_n}  =  \sqrt{12 \pi^2 f^3 \, ({\cal N} / {\cal T}) }  \simeq   10^{-14} \times \nonumber\\ 
%&\!\!\!\!\!\!  \!\!\!\!\!\! \!\!\!\!\!\! \!\!\!\!\!\! \!\!\!\!\!\!  \!\!\!\!\!\! \!\!\!\!\!\!  \simeq& \!\!\!  \!\!\!\!\!\!  \!\!\!\!\!\! \!\!\!\!\!\!   10^{-14} 
 &\!\!\!\!\!\!  \!\!\!\!\!\! \!\!\!\!\!\! \!\!\!\!\!\! \!\!\!\!\!\!  \!\!\!\!\!\! \!\!\!\!\!\!  \times& \!\!\!  \!\!\!\!\!\! \!\!\!\!\!\!  \!\!\!  \!\!\!\!\!\! \;\sqrt{ \frac{ \Delta t_{14d} \;  \sigma^2_{\mu s}}{ {T_{\rm obs, 15yr}^{3}}}} \;    \left( \xi_{\rm frac} \cdot \left( f \cdot T_{\rm obs} \right)^{-3/2} + \left( f \cdot T_{\rm obs} \right)^{3/2} \right) % \nonumber \!\!\!\!\!\!  \!\!\!\!\!\! \!\!\!\!\!\!  \\ \!\!\!\!\!\! 
\label{eqnoisecurve}
\end{eqnarray}
where  $T_{\rm obs, 15yr}=T_{\rm obs}/15{\rm yr}$ is scaled observation time, $\Delta t_{14d}= \frac{\Delta t}{14{\rm days}}$ is cadence, $ \sigma_{\mu s}= \frac{ \sigma}{ {\rm \mu sec}}$ is rms signal error ,  $\xi_{\rm frac}$ is the inverse of the square-root of the fraction of pulsars where red noise is subdominant compared to white noise. We take 1/4th of the pulsars to be white-noise dominated in the regime we focus on, so $\xi_{frac}=2$ in this study. We emphasize that Eq.~\eqref{eqnoisecurve} is a generic result for PTAs, hence plugging in distinct timing errors or observation periods, one can produce approximate sensitivity curves. The scalings correspond to current PTA experiments, namely the rms timing error is normalized for ${\rm \mu s}$. The next generation SKA experiment is expected to improve this result by an order of magnitude, i.e.\ to $30\,{\rm ns}$. Moreover, the number of observed pulsars with SKA will be about two orders of magnitude larger than the current PTA pulsar number \cite{Smits:2008cf}. Combined, these two effects will improve the sensitivity of SKA by roughly two orders of magnitude, as shown in Fig. \ref{signalvssensitivity}.

To proceed, we plug Eq.~\eqref{eqsignal} into Eq.~\eqref{eq:snrsquare}, using our source is monochromatic, with $f_*$  given in Eq.~\eqref{eqfreqmass}
\begin{eqnarray}
%{\rm SNR}^2 &\simeq& 2\, T_{\rm obs} \int df  \; {\cal R}^2_{\beta} \; \left( \frac{h_{c,{\rm signal}}}{h_{c,{\rm noise}}} \right)^4 \nonumber\\
 {\rm SNR}^2 &\simeq& 2\, ( f_* \cdot T_{\rm obs})^2   \;  {\cal R}^2_{\beta} \;   \left( \frac{h_{c,{\rm signal}}}{h_{c,{\rm noise}}} \right)^4 \bigg|_{f=f_*} \; ,
 \label{eqlabelsnr2}
\end{eqnarray}
where  ${\cal R}^2_{\beta}$ is the response given to different spin fields, i.e.\ scalar, vector and tensor, given in Eq.~\eqref{eqptaspinresponse}.  With this result in hand, we are  equipped with all the required items to compute the sensitivity of  current and future PTA experiments to large portions of the parameter space of ULDM particles of each distinct nature (scalar, vector and tensor).
Our results are shown in Figs.~\ref{figbigpicture}, and~\ref{figscalarvectorspin2zoom}. 

To provide intuition for the results, we now  discuss in  detail the capabilities of PTAs to probe ULDM by focusing on the scaling properties of the signal and noise in each separate frequency regime in  Figs.~\ref{signalvssensitivity},~\ref{figbigpicture} and~\ref{figscalarvectorspin2zoom} by assuming conservative noise curve.
\begin{itemize}[leftmargin=*]
\item[]{a)}
$f>1/T_{\rm obs}$ : For scalar and vectors we have $h_{\rm signal}\propto f^{-2}$, while for tensor(spin-2) $h_{\rm signal}\propto f^{-1}$; and $h_{\rm noise}\propto f^{3/2}$.
\begin{eqnarray}
 S,V &:& \; {\rm SNR}^2 \propto  (f\cdot T_{obs})^2\,  \frac{h_{c,s}^4}{h_{c,n}^4} \propto {\cal F}_{\rm ULDM}^{\, 4} \; (f \cdot T_{\rm obs})^{-12} \nonumber\\
 T &:&\;   {\rm SNR}^2 \propto (f\cdot T_{obs})^2 \,   \frac{h_{c,s}^4}{h_{c,n}^4}  \propto {\cal F}_{\rm ULDM}^{\, 2} \; (f \cdot T_{\rm obs})^{-8}  \;\;\;\;  \;\;\; \;\;\;
\end{eqnarray} 
The fraction of dark matter we can probe scales as ${\cal F}_{\rm ULDM}\propto (f\cdot T_{\rm obs})^{3}$ for scalar and vector, and ${\cal F}_{\rm ULDM}\propto (f\cdot T_{\rm obs})^{4}$ for tensor. This is the regime for masses larger than $10^{-23}\,{\rm eV}$. %(assuming $T_{\rm obs}=15\,{\rm yrs}$).

\item[]{b)}
$1/D_{\rm pulsar}<f<1/T_{\rm obs}$: For scalar and vectors we have $h_{\rm signal}\propto f^{-2}$, while for tensor $h_{\rm signal}\propto f^{-1}$; and $h_{\rm noise}\propto f^{-3/2}$.
\begin{eqnarray}
 S,V &:& \;  {\rm SNR}^2 \propto (f\cdot T_{obs})^2 \,  \frac{h_{c,s}^4}{h_{c,n}^4} \propto {\cal F}_{\rm ULDM}^{\, 4} \; (f \cdot T_{\rm obs})^{0} \nonumber\\
 T &:&\;   {\rm SNR}^2  \propto (f\cdot T_{obs})^2 \,  \frac{h_{c,s}^4}{h_{c,n}^4} \propto {\cal F}_{\rm ULDM}^{\, 2} \; (f \cdot T_{\rm obs})^{4}  \;\;\;\;  \;\;\; \;\;\;
\end{eqnarray} 
For scalars and vectors, in the $f<1/T_{\rm obs}$ regime, the signal grows slightly faster than noise, which results in ${\cal F}_{\rm ULDM}\propto (f\cdot T_{\rm obs})^{0}$. In the tensor case, we have ${\cal F}_{\rm ULDM}\propto (f\cdot T_{\rm obs})^{-2}$. Depending on $T_{\rm obs}$ of the PTA, this regime  typically lies  between $m\sim10^{-26}\,{\rm eV}$ and $m\sim10^{-23}\,{\rm eV}$.

\item[]{c)}
$f<1/D_{\rm pulsar}$: For scalar and vectors $h_{\rm signal}\propto f^{-1}$, while for tensor $h_{\rm signal}\propto f^{0}$; and $h_{\rm noise}\propto f^{-3/2}$.
\begin{eqnarray}
S,V &:& \;  {\rm SNR}^2 \propto (f\cdot T_{obs})^2 \,  \frac{h_{c,s}^4}{h_{c,n}^4} \propto {\cal F}_{\rm ULDM}^{\, 4} \; (f \cdot D_{\rm pulsar})^{4} \nonumber\\
 T &:&\;  {\rm SNR}^2  \propto (f\cdot T_{obs})^2 \,  \frac{h_{c,s}^4}{h_{c,n}^4} \propto {\cal F}_{\rm ULDM}^{\, 2} \; (f \cdot D_{\rm pulsar})^{8} \;\;\;\;  \;\;\; \;
\end{eqnarray}
In the regime $f<1/D_{pulsar}$, the ${\rm SNR}$ decreases for scalar, vector and tensor. The fraction of dark matter that can be probed scales as  ${\cal F}_{\rm ULDM}\propto (f \cdot D_\pls)^{-1}$ for scalar and vector; and ${\cal F}_{\rm ULDM}\propto (f \cdot D_\pls)^{-4}$ for tensor.
Depending on the pulsar distance $D_\pls$, this is typically the regime where $m\lesssim10^{-26}$eV.
\end{itemize}

To conclude, our results indicate that current as well as future PTAs can probe ultralight bosons of scalar, vector (dark photon) and tensor (spin-2) types with excellent precision.  
 The mass range $10^{-26}-10^{-23}\,{\rm eV}$ is especially interesting since CMB and large-scale structure experiments have less sensitivity in that regime, and in contrast PTAs have the most sensitivity there, so that we can close this gap, as shown in Fig.~\ref{figbigpicture} and~\ref{figscalarvectorspin2zoom}. Crucially, this mass range roughly corresponds 
to the frequency regime $1/D_{\rm pulsar}\!<\!f\!<\!1/T_{\rm obs}$, where the signal stays strong compared to the noise. Therefore, smaller mass particles (in the scalar and vector scenarios) can be tightly constrained until frequencies $f\!\sim\!1/D_{\rm pulsar}$. For tensor particles, the potential oscillations scale with $1/f$, hence the best regime to probe them is around  $f\!\sim\! 1/T_{\rm obs}$.
 
 Our calculations show that with current PTA data, the abundance of ultralight bosons in the mass range $10^{-26}-10^{-23}\,{\rm eV}$ can be probed down to ${\cal O} (1-10)\%$ of the total dark matter energy density. We also found that with 30 year PTA data, the precision improves by about one order of magnitude to $1 \%$, and with SKA to $0.1\%$.

This work implies that combining PTAs with current constraints from CMB, large-scale structure, Lyman-$\alpha$ and superradiance (see Fig.~\ref{figbigpicture}), ultralight scalar dark matter can be constrained throughout the mass range $10^{-30}\!-\!10^{-17}\,{\rm eV}$ to less than $\mathcal{O}(10\%)$  of the  dark matter.

\vspace{0.15in}

{\it Acknowledgements.}
We thank 
Jeff Hazboun, Sarah Libanore, David Marsh, Andrea Mitridate, Joseph Romano, Debanjan Sarkar, Marc Kamionkowski, Michael Lam, Sarah Vigeland, Stephen Taylor for  discussions and  feedback on the manuscript, and Jordan Flitter for the SPARC bounds. We thank especially Tristan Smith for the detailed explanations on low frequency signal, sensitivity and feedback on the draft. We also thank the anonymous referees for suggesting important corrections and improvements. C\"U thanks his family for their support during this work, and dedicates this work to Ece Ceyda G\"udemek and R\"umeysa Berin \c{S}en. C\"U is supported by the Kreitman fellowship of BGU, and the Excellence fellowship of the Israeli Academy of Sciences and Humanities, and the Council for Higher Education.  FU is supported by the European Regional Development Fund (ESIF/ERDF) and the Czech Ministry of Education, Youth and Sports (MEYS) through Project CoGraDS-\verb|CZ.02.1.01/0.0/0.0/15_003/0000437|. EDK is supported by an  Azrieli Foundation Faculty Fellowship.

\end{document}